\pdfoutput=1
\documentclass[aps,prd,amsmath,floats,floatfix, twocolumn,
superscriptaddress,nofootinbib,showpacs,longbibliography]{revtex4-1}
\usepackage[T1]{fontenc}
\usepackage[utf8]{inputenc}
\usepackage{lmodern}
\usepackage{verbatim}
\usepackage[dvipsnames, usenames]{xcolor}
\definecolor{linkcolor}{rgb}{0.0,0.3,0.5}
\usepackage[hypertexnames=false, unicode, colorlinks=true, linkcolor=linkcolor,
citecolor=linkcolor, filecolor=linkcolor,urlcolor=linkcolor,
pdfusetitle]{hyperref}
\usepackage[all]{hypcap}
\usepackage{graphicx}
\usepackage{xspace}
\usepackage{amssymb}
\usepackage{microtype}

\usepackage[english]{babel}
\usepackage{blindtext}

\usepackage[normalem]{ulem} 
\usepackage{bm} 
\usepackage{mathrsfs}

\graphicspath{
	{plots/}
}

\DeclareMathAlphabet{\mathpzc}{OT1}{pzc}{m}{it}
\newcommand{\tousif}[1]{\textcolor{blue}{#1}}
\usepackage[nomessages]{fp}

\begin{document}
\title{On the approximate relation between black-hole perturbation theory \\and numerical relativity}
\newcommand{\UMassDMath}{\affiliation{Department of Mathematics,
		University of Massachusetts, Dartmouth, MA 02747, USA}}
\newcommand{\UMassDPhy}{\affiliation{Department of Physics,
		University of Massachusetts, Dartmouth, MA 02747, USA}}
\newcommand{\CSCVRUMass}{\affiliation{Center for Scientific Computing and Data Science Research, University of Massachusetts, Dartmouth, MA 02747, USA}}
\newcommand{\URI}{\affiliation{Department of Physics and Center for Computational Research, 
    University of Rhode Island, Kingston, RI 02881, USA}}    

\author{Tousif Islam}
\email{tislam@umassd.edu}
\UMassDPhy
\UMassDMath
\CSCVRUMass

\author{Gaurav Khanna}
\URI
\UMassDPhy
\CSCVRUMass

\hypersetup{pdfauthor={Islam et al.}}

\date{\today}

\begin{abstract}
We investigate the interplay between numerical relativity (NR) and adiabatic point-particle black hole perturbation theory (ppBHPT) in the comparable mass regime for quasi-circular non-spinning binary black holes. Specifically, we reassess the $\alpha$-$\beta$ scaling technique, previously introduced by Islam \textit{et al.}~\cite{Islam:2022laz}, as a means to effectively match ppBHPT waveforms to NR waveforms within this regime. In particular, $\alpha$ rescales the amplitude and $\beta$ rescales the time (and hence the phase). Utilizing publicly available long NR data (\texttt{SXS:BBH:2265}~\cite{sxs_collaboration_2019}) for a mass ratio of $1:3$, encompassing the final $\sim 65$ orbital cycles of the binary evolution, we examine the range of applicability of such scalings. We observe that the scaling technique remains effective even during the earlier stages of the inspiral. Additionally, we provide commentary on the temporal evolution of the $\alpha$ and $\beta$ parameters and discuss whether they can be approximated as constant values. Consequently, we derive the $\alpha$-$\beta$ scaling as a function of orbital frequencies and demonstrate that it is equivalent to a frequency-dependent correction. We further provide a brief comparison between post-Newtonian (PN) waveforms and the rescaled ppBHPT waveform at a mass ratio of $1:3$ and comment on their regime of validity. Finally, we explore the possibility of using PN theory to obtain the $\alpha$-$\beta$ calibration parameters and still provide a rescaled ppBHPT waveform that matches NR.
\end{abstract}

\maketitle

\begin{figure}
\includegraphics[width=\columnwidth]{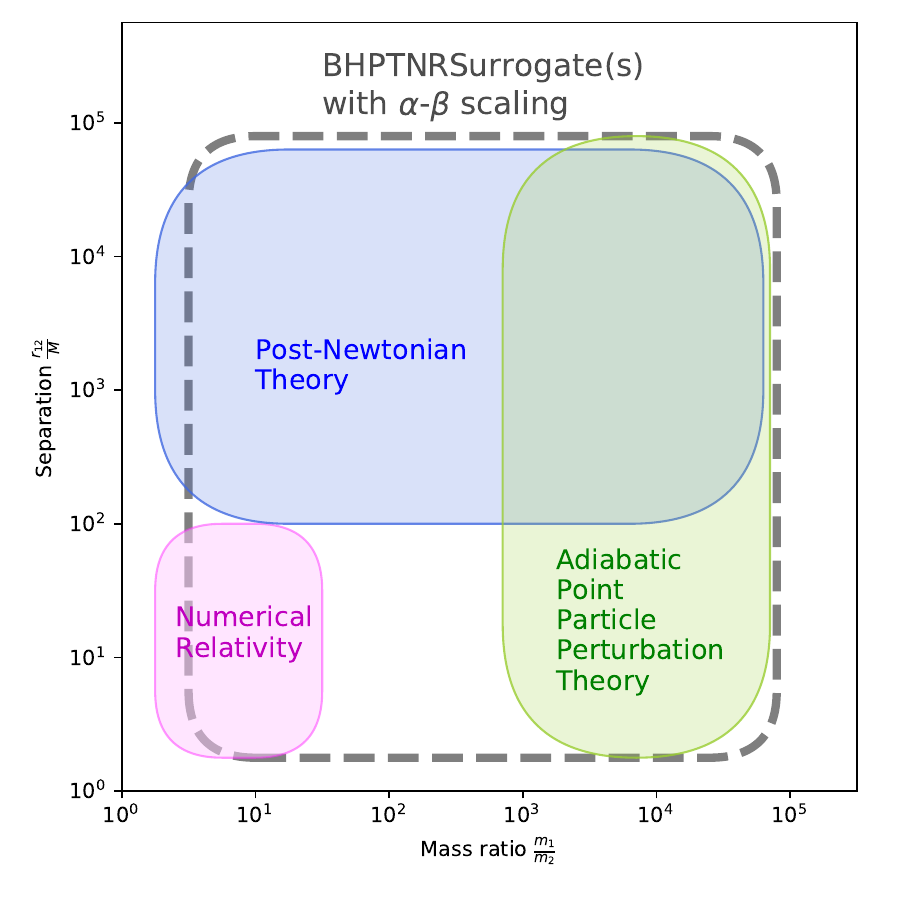}
\caption{We show the approximate schematic of the regime of validity of three different methods to simulate BBH mergers: NR, ppBHPT and post-Newtonian approximations. For comparison, we show the regime of validity of the NR-tuned ppBHPT surrogate models (named collectively as \texttt{BHPTNRSurrogate(s)}) that employs a rescaling technique called $\alpha$-$\beta$ scaling to match ppBHPT waveforms to NR in the comparable mass regime. Here, x-axis shows the mass ratio $\frac{m_1}{m_2}$ whereas y-axis shows the separation $r_{12}$ (scaled by the total mass of the binary $M:=m_1+m_2$) between the two component black holes in a binary. More details are in Section \ref{Sec:Introduction}.}
\label{fig:regime}
\end{figure}

\section{Introduction}
\label{Sec:Introduction}
Development of computationally efficient yet precise waveform models~\cite{Blackman:2015pia,Blackman:2017pcm,Blackman:2017dfb,Varma:2018mmi,Varma:2019csw,Islam:2021mha,bohe2017improved,cotesta2018enriching,cotesta2020frequency,pan2014inspiral,babak2017validating,husa2016frequency,khan2016frequency,london2018first,khan2019phenomenological} binary-black-hole (BBH) mergers play a crucial role in GW research.
This relies heavily on accurate numerical simulations of BBH mergers. In the regime of comparable mass ratios ($1 \le q \le 10$ where $q:=m_1/m_2$ represents the mass ratio of the binary, with $m_1$ and $m_2$ denoting the masses of the primary/larger and secondary/smaller black holes, respectively), the most accurate approach to simulate a BBH merger is by solving the Einstein equations using numerical relativity (NR)~\cite{Mroue:2013xna,Boyle:2019kee,Healy:2017psd,Healy:2019jyf,Healy:2020vre,Healy:2022wdn,Jani:2016wkt,Hamilton:2023qkv} (Fig.~\ref{fig:regime}). However, accurately simulating BBH mergers using NR in the intermediate to large mass ratio regime ($10 \leq q \leq 100$) remains a challenging task due to algorithmic complexity. 

In contrast, adiabatic point particle black hole perturbation theory (ppBHPT)~\cite{Sundararajan:2007jg,Sundararajan:2008zm,Sundararajan:2010sr,Zenginoglu:2011zz,Fujita:2004rb,Fujita:2005kng,Mano:1996vt,throwe2010high,OSullivan:2014ywd,Drasco:2005kz} offers a reliable modeling approach for extreme mass ratio inspirals (EMRI) ($q \to \infty$) (Fig.~\ref{fig:regime}). In ppBHPT, the smaller black hole is treated as a point particle orbiting the larger black hole described by a curved space-time background. However, as the binary system becomes less asymmetric and approaches the regime of comparable mass ratios, the assumptions of the ppBHPT framework begin to break down. Consequently, the ppBHPT framework fails to generate accurate gravitational waveforms within this regime. On the other hand, post-Newtonian (PN) theories~\cite{Blanchet:2013haa} provide a dependable approximate method to generate gravitational waveforms for BBH mergers during the inspiral stage of the binary evolution when the two black holes are considerably distant from each other and their velocities are significantly smaller than the speed of light (Fig.~\ref{fig:regime}).

In recent times, there have been significant advancements in expanding the scope of both NR and ppBHPT frameworks. These advancements include the development of the \texttt{BHPTNRSur1dq1e4} surrogate model \cite{Islam:2022laz,Rifat:2019ltp}, a fully relativistic second-order self-force model \cite{Wardell:2021fyy}, and the extension of NR techniques to simulate BBH mergers with higher mass ratios \cite{Lousto:2020tnb,Lousto:2022hoq,Yoo:2022erv,Giesler:2022inPrep}. 

The \texttt{BHPTNRSur1dq1e4} surrogate model, which relies on the ppBHPT framework, has exhibited reasonable accuracy in predicting waveforms for BBH mergers in the comparable to large mass ratios regime. By employing a straightforward calibration procedure known as the $\alpha$-$\beta$ scaling, the ppBHPT waveforms are appropriately rescaled to achieve excellent match with NR data, particularly in the comparable mass regime.
The scaling reads~\cite{Islam:2022laz}:
\begin{align} \label{eq:EMRI_rescale}
h^{\ell,m}_{\tt NR}(t_{\tt NR} ; q) \sim {\alpha_{\ell}} h^{\ell,m}_{\tt ppBHPT}\left( \beta  t_{\tt ppBHPT};q \right) \,,
\end{align}
where, $h^{\ell,m}_{\tt NR}$ and $h^{\ell,m}_{\tt ppBHPT}$ represent the NR and ppBHPT waveforms, respectively, as functions of the NR time $t_{\tt NR}$ and ppBHPT time $t_{\tt ppBHPT}$. The calibration parameters, $\alpha_{\ell}$ and $\beta$, are typically determined through matching ppBHPT  waveforms to NR. 
Following the $\alpha$-$\beta$ calibration procedure, the quadrupolar mode of the rescaled ppBHPT waveform exhibits an excellent agreement with NR, with errors of approximately $10^{-3}$ or less, in the comparable mass regime~\cite{Islam:2022laz}. Additionally, the rescaled ppBHPT waveforms demonstrate a remarkable match to recently obtained NR data in the high mass ratio regime ($q=15$ to $q=128$)~\cite{Islam:2023qyt}. It has been shown that these waveforms can be used to accurately estimate the properties of the final black holes~\cite{Islam:2023mob}. Further analysis provides evidence that the calibration parameters can be attributed to the absence of finite size effects within the ppBHPT framework~\cite{Islam:2023aec}.

In this paper, we investigate the interplay between NR and ppBHPT in the comparable mass regime through the lens of the $\alpha$-$\beta$ scaling. In particular, we use publicly available long NR data (\texttt{SXS:BBH:2265}), for \tousif{a} mass ratio of $q=3$, that covers the final $\sim 65$ orbital cycles of the binary evolution to understand applicability of the $\alpha$-$\beta$ scaling. This particular NR data is almost ten times longer in duration than most of the existing NR data and have significantly more cycles. In Section~\ref{Sec:Scaling}, we present our main findings and results. To begin, Section~\ref{sec:alpha_beta_methods} explores different methods to obtain the calibration parameters $\alpha$ and $\beta$. Next, in Section~\ref{sec:alpha_beta_comparison}, we provide a detailed comparison of the $\alpha$ and $\beta$ values obtained from these different approaches. Section~\ref{sec:alpha_beta_validity} then comments on the regime of validity of the $\alpha$-$\beta$ scaling. Finally, in Section~\ref{Sec:Discussion}, we discuss the implication of our results in current and future efforts in modeling gravitational waveforms from BBH mergers.

\section{Scaling between NR and perturbation theory}
\label{Sec:Scaling}
In this section, we present a detailed analysis of the $\alpha$-$\beta$ scaling between ppBHPT and NR waveforms in the comparable mass regime. To do this, we utilize publicly available long NR data (\texttt{SXS:BBH:2265}~\cite{sxs_collaboration_2019}), for a mass ratio of $q=3$. The NR data covers the final $\sim 65$ orbital cycles of the binary evolution and are $\sim 30000M$ long in duration (where $M$ is the total mass of the binary). We then generate the ppBHPT waveform for this mass ratio using the framework developed in Refs.~\cite{Sundararajan:2007jg,Sundararajan:2008zm,Sundararajan:2010sr,Zenginoglu:2011zz}. In particular, we first compute the full inspiral-merger-ringdown trajectory taken by the point-particle and then we use that trajectory to compute the gravitational wave emission by solving the inhomogeneous Teukolsky equation in the time-domain \cite{Sundararajan:2007jg,Sundararajan:2008zm,Sundararajan:2010sr,Zenginoglu:2011zz,Field:2021}. A brief summary of our framework is given in Section II of Ref.~\cite{Islam:2022laz}. Our ppBHPT waveform data covers the final $\sim 56$ orbital cycles of the binary evolution and are $\sim 35000m_1$ long in duration. 
\subsection{Methods to obtain $\alpha$-$\beta$ values}
\label{sec:alpha_beta_methods}
Once we have both the ppBHPT and NR data for $q=3$, we investigate various methods to determine the appropriate $\alpha_\ell$ and $\beta$ values necessary for accurately rescaling the ppBHPT waveform to achieve a strong agreement with NR. To simplify the analysis, we focus on the $(2,2)$ mode of the waveform. Subsequently, we drop the subscript $\ell$ and use only $\alpha$ unless otherwise mentioned.

\begin{figure*}
\includegraphics[width=\textwidth]{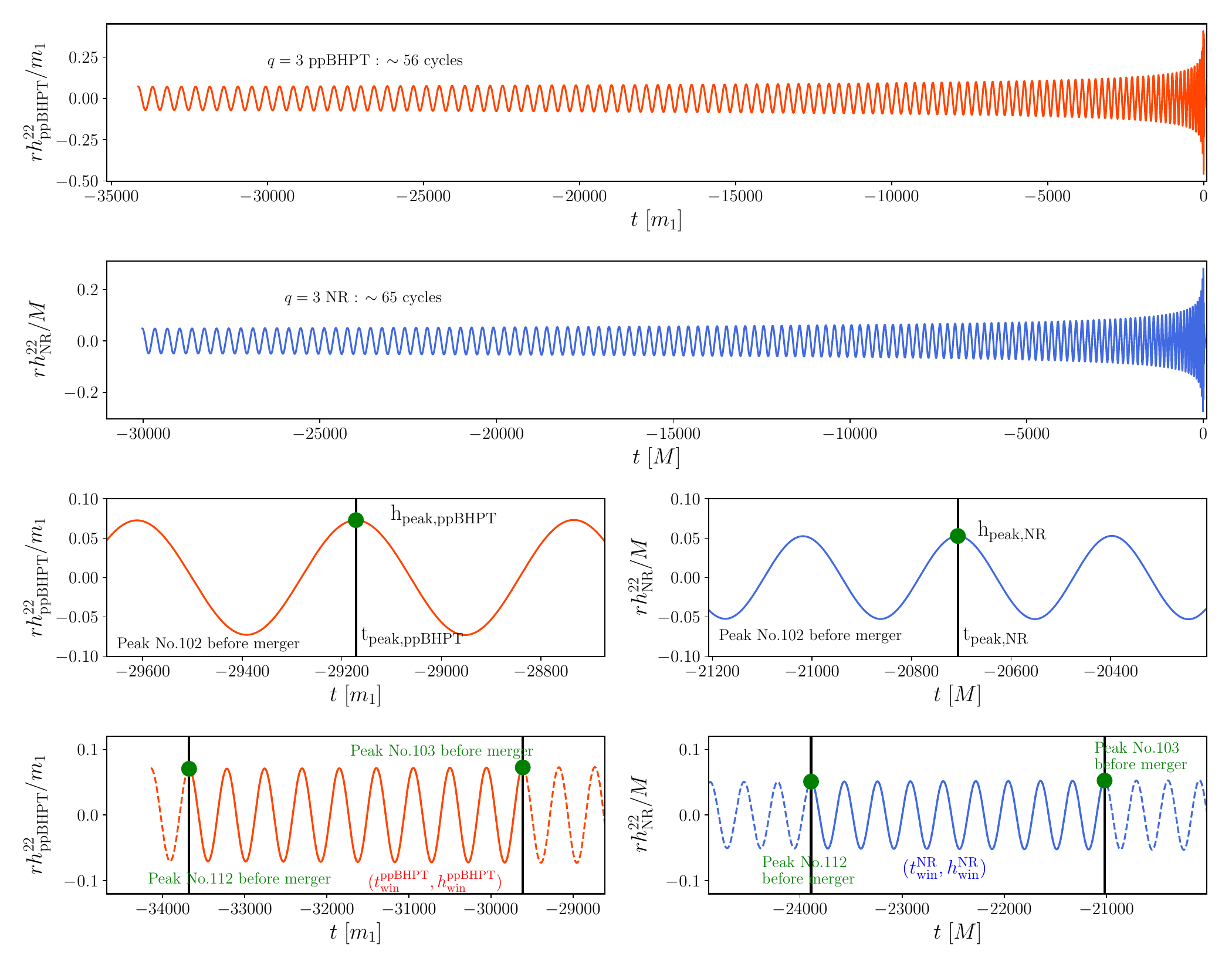}
\caption{Comparison of the $(2,2)$ mode of the NR (first row) and ppBHPT (second row) waveforms for $q=3$, along with the $102^{\rm nd}$ peak of both waveforms (third row) and the waveform segment between the $102^{\rm nd}$ and $103^{\rm rd}$ peaks (fourth row). This visualization demonstrates the waveform portion used in different approaches to estimate the $\alpha$ and $\beta$ values. More details are in Section \ref{sec:alpha_beta_comparison}.}
\label{fig:q3_beta_peaks}
\end{figure*}

\subsubsection{Using full waveform data}
\label{sec:scaling_full_wf}
Typically, the values of $\alpha$ and $\beta$ are determined by minimizing the $L_2$-norm difference between the NR data and the rescaled ppBHPT waveforms, covering full inspiral-merger-ringdown stage of the binary evolution, after aligning them on the same time grid~\cite{Islam:2022laz,Rifat:2019ltp}. The optimization problem can be formulated as follows:
\begin{align} \label{eq:alpha_lm}
\min_{\alpha,\beta} \frac{\int \left| \alpha h^{2,2}_{\tt ppBHPT}\left( \beta t_{\tt ppBHPT} \right) - h^{2,2}_{\tt NR}(t_{\tt NR}) \right|^2 dt}{\int \left| h^{2,2}_{\tt NR}(t_{\tt NR}) \right|^2 dt}.
\end{align}
This optimization problem yields the global best-fit values of $\alpha$ and $\beta$ that minimize the error computed over the entire length of the waveform data or the calibration regime (e.g. $t \in [-5000,100]M$ as used in ~\cite{Islam:2022laz}).

\subsubsection{Using only inspiral data}
\label{sec:scaling_inspiral_wf}
We can modify the procedure described in Section \ref{sec:scaling_full_wf} by limiting the global fit to only include inspiral data, such as data up to $t=-100M$. This approach eliminates the influence of the merger-ringdown portion of the waveform, which may have different mass scale and spin values.

\subsubsection{Using the peaks}
\label{sec:scaling_peak}
Alternatively, it is possible to estimate the optimal values of $\alpha$ and $\beta$ at different points during the binary evolution. However, special care should be taken as this approach requires simultaneous rescaling of both the time and amplitude. 

We note that, in order to achieve a successful rescaling, it is necessary for the peaks of the waveform to align between the NR and ppBHPT data for a given cycle before merger. Therefore, we can estimate the optimal values of $\alpha$ and $\beta$ at each peak by matching the peak time and value between NR and ppBHPT. For instance, we can focus on the $50^{\rm th}$ peak before the merger in both NR and ppBHPT waveforms. By employing cubic splines, we can accurately determine the precise location and value of the peak from the discrete waveform data in both cases. Let us denote the peak times as $t_{\rm peak, ppBHPT}$ and $t_{\rm peak, NR}$, while the peak values are denoted as $h_{\rm peak, ppBHPT}$ and $h_{\rm peak, NR}$. In this analysis, the point estimates of $\alpha$ and $\beta$ at the peaks are given by:
\begin{equation}
\alpha_{\rm peak} = \frac{h_{\rm peak, NR}}{h_{\rm peak, ppBHPT}},
\end{equation}
and
\begin{equation}
\beta_{\rm peak} = \frac{t_{\rm peak, NR}}{t_{\rm peak, ppBHPT}}.
\end{equation}
By repeating this analysis for all the peaks, we can obtain a temporal variation of the optimal local values of $\alpha$ and $\beta$ throughout the binary evolution.

\subsubsection{Using a certain number of cycles}
\label{sec:scaling_window}
Finally, we can modify the method to estimate the local values of $\alpha$ and $\beta$ throughout the binary evolution by considering a broader time window instead of just focusing on individual peaks. For example, we can choose to match the ppBHPT and NR waveforms between the 50th and 41st peak before the merger. The shorter duration NR and ppBHPT data, which are restricted to the selected time window, can be denoted as $(t^{\rm NR}{\rm win},h^{\rm NR}{\rm win})$ and $(t^{\rm ppBHPT}{\rm win},h^{\rm ppBHPT}{\rm win})$, respectively. We then perform the $\alpha$-$\beta$ scaling as described in Eq. (\ref{eq:EMRI_rescale}) on these dataset by minimizing the following difference:
\begin{align} \label{eq:alpha_lm}
\min_{\alpha,\beta} \frac{\int \left| \alpha h_{\tt ppBHPT}^{\rm win}\left(\beta t_{\tt ppBHPT}^{\rm win} \right) - h_{\tt NR}^{\rm win}(t_{\tt NR}^{\rm win}) \right|^2 dt}{\int \left| h_{\tt NR}^{\rm win}(t_{\tt NR}^{\rm win}) \right|^2 dt}.
\end{align}
This minimization problem yields the global best-fit values of $\alpha$ and $\beta$ that minimize the error computed over the entire length of the waveform data or the calibration regime (e.g. $t \in [-5000,100]M$ as used in ~\cite{Islam:2022laz}).

This approach allows us to obtain an averaged local estimate of the $\alpha$ and $\beta$ values around the time corresponding to the mean of the time window between the 50th and 41st peak before the merger. In this modified approach, we utilize 10 cycles of waveform data to estimate the $\alpha$ and $\beta$ values, which we denote as $\alpha_{\rm 5cycles}$ and $\beta_{\rm 5cycles}$.  

\begin{figure}
\includegraphics[width=\columnwidth]{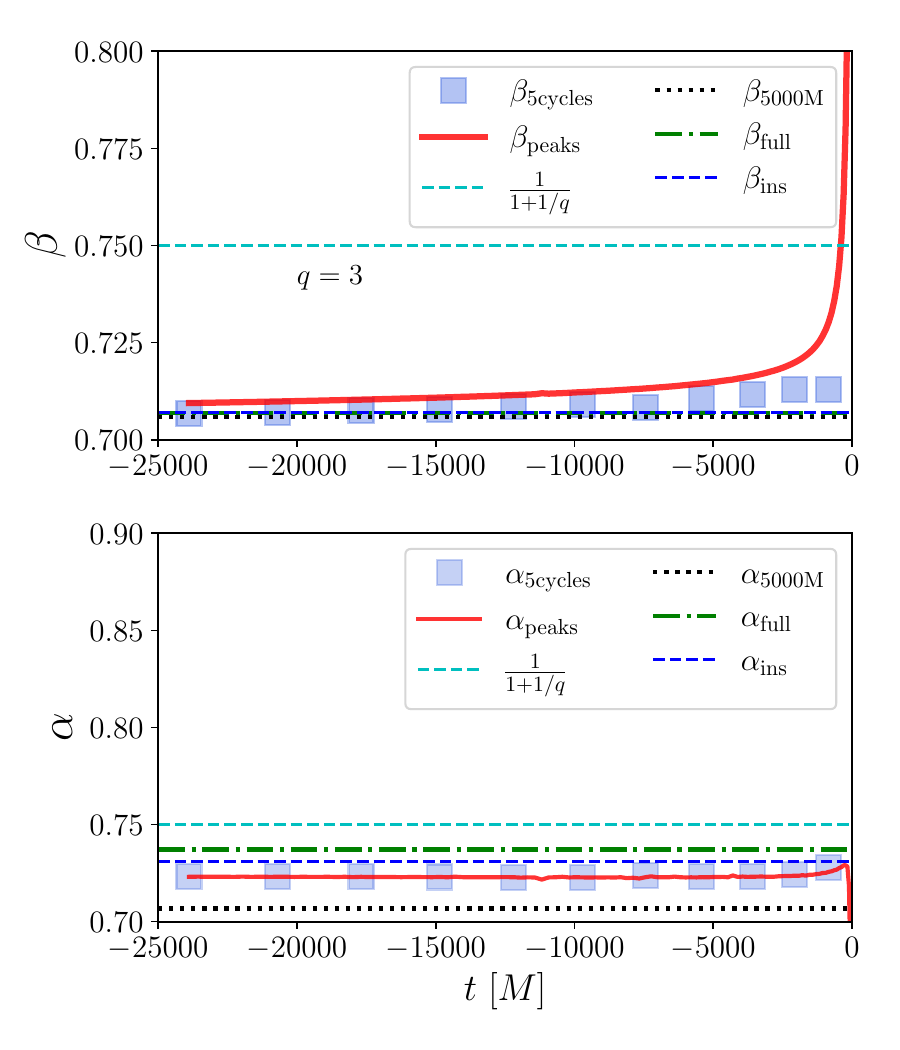}
\caption{We show the $\alpha$ and $\beta$ values for $q=3$ obtained from different approaches outlined in Section~\ref{sec:alpha_beta_methods}. For comparison, we also show the naive scaling of $\frac{1}{1+1/q}$ required to ensure consistency in mass-scale between ppBHPT and NR. More details are in Section \ref{sec:alpha_beta_comparison}.}
\label{fig:q3_optimal_alpha_beta}
\end{figure}

\subsection{Comparison of the $\alpha$-$\beta$ values from different methods}
\label{sec:alpha_beta_comparison}
To infer both global and local estimates of the $\alpha$ and $\beta$ values for $q=3$, we first employ three different techniques:
\begin{itemize}
    \item We use the final $\sim 5000M$ of the NR data to find the global best-fit values of $\alpha$ and $\beta$. This is done by minimizing the $L_2$-norm difference between the rescaled ppBHPT waveform and the NR waveform, as described in Section~\ref{sec:scaling_full_wf}. The obtained calibration values are denoted as $\alpha_{5000M}$ and $\beta_{5000M}$.
    \item We match all 112 peaks in the NR data to their corresponding peaks in the ppBHPT waveform using the procedure outlined in Section~\ref{sec:scaling_peak}. This gives us the point estimates of $\alpha$ and $\beta$ at each peak, denoted as $\alpha_{\rm peak}$ and $\beta_{\rm peak}$.
    \item The NR data is divided into smaller windows consisting of 10 consecutive peaks (i.e. 5 cycles), resulting in 10 smaller time windows. We then apply the procedure described in Section~\ref{sec:scaling_window} to match each of these smaller windows to the corresponding ppBHPT waveforms. This provides us with the averaged local estimations of the calibration parameters, denoted as $\alpha_{\rm 5cycles}$ and $\beta_{\rm 5cycles}$.
\end{itemize}
By employing these three techniques, we can obtain a comprehensive understanding of the $\alpha$ and $\beta$ values for the considered mass ratio of $q=3$.

\begin{figure}
\includegraphics[width=\columnwidth]{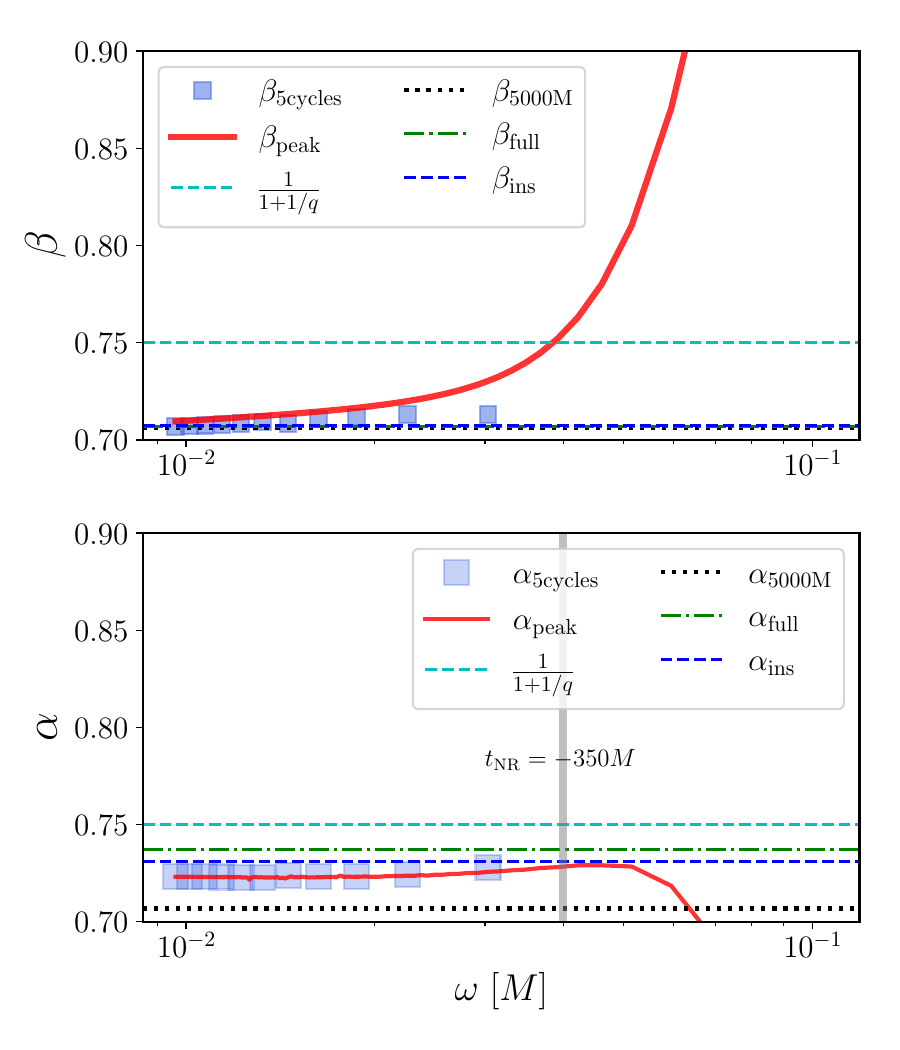}
\caption{We show the $\alpha$ and $\beta$ values for $q=3$ obtained from different approaches as a function of the NR orbital frequencies. For comparison, we also show the naive scaling of $\frac{1}{1+1/q}$ required to ensure consistency in mass-scale between ppBHPT and NR. More details are in Section \ref{sec:alpha_beta_comparison}.}
\label{fig:q3_optimal_alpha_beta_freq}
\end{figure}

\begin{figure*}
\includegraphics[width=\textwidth]{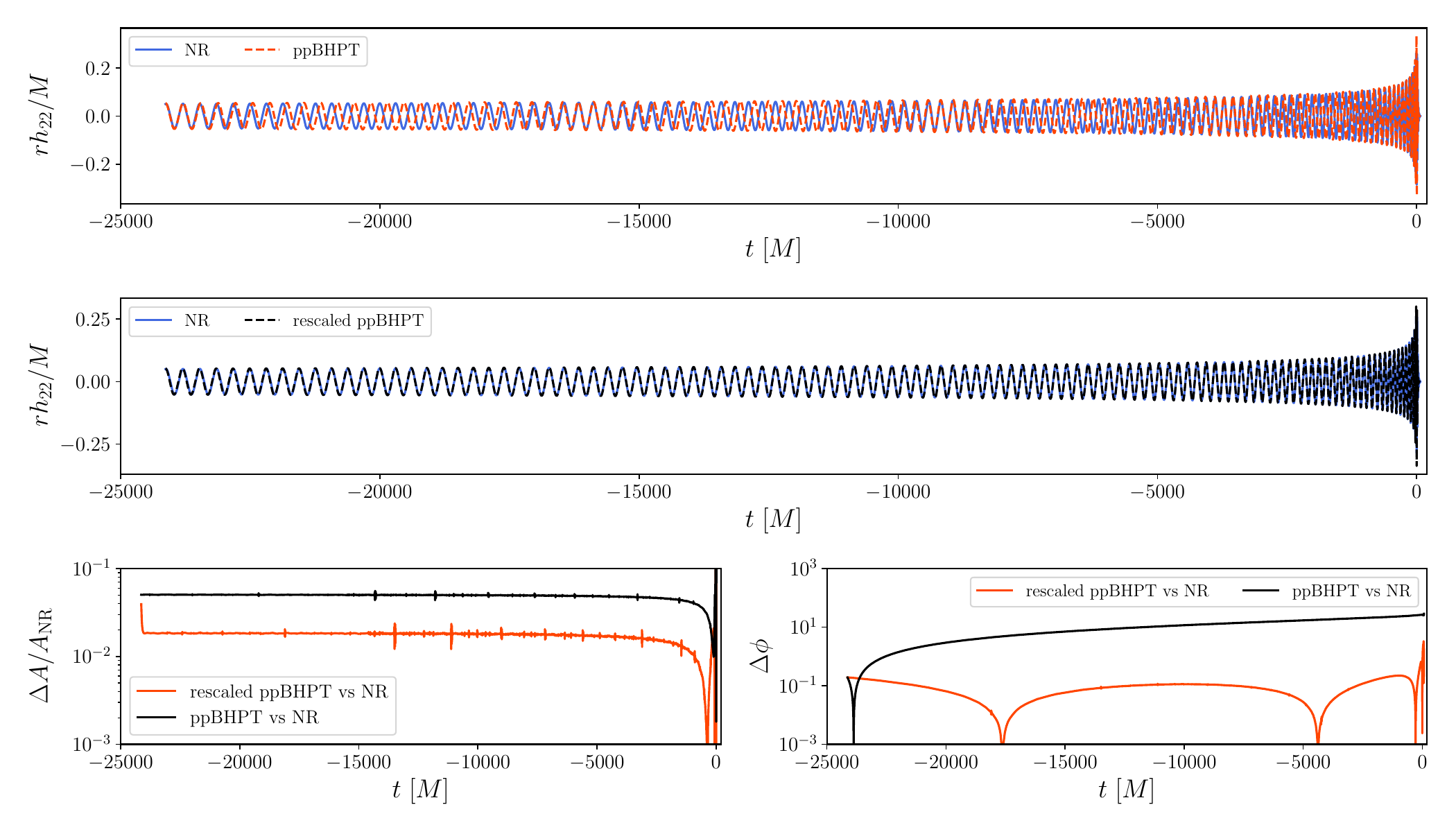}
\caption{We show the comparison of the $(2,2)$ mode of the NR  and ppBHPT waveforms (first row), along with the comparison between NR and rescaled ppBHPT waveforms (second row) for $q=3$. Additionally, we show the errors in the amplitudes and phases (third rows) for both the waveforms when compared to NR. All waveforms have the mass-scale of $M$. More details are in Section \ref{sec:alpha_beta_comparison}.}
\label{fig:q3_alpha_beta_optimization}
\end{figure*}

Figure~\ref{fig:q3_beta_peaks} illustrates the $(2,2)$ mode of the NR (first row) and ppBHPT waveforms (second row), both aligned such that the maximum amplitude occurs at $t=0$ and the orbital phase is zero at the beginning. This alignment facilitates a direct comparison between the two waveforms. Additionally, we highlight the $102^{\rm nd}$ peak of both waveforms (third row), along with their corresponding peak times. It is evident that the peak times and values differ between the NR and ppBHPT waveforms due to dephasing between the NR and adiabatic approximation of the ppBHPT. The peaks in the ppBHPT waveform occur earlier in time and have larger amplitudes compared to the NR waveform. This emphasizes the need to establish a scaling relationship between the ppBHPT and NR waveforms. Finally, we show the waveform segment between the $102^{\rm nd}$ and $103^{\rm rd}$ peaks for both NR and ppBHPT as a demonstration of the procedure mentioned in Section~\ref{sec:scaling_window}.

\begin{figure*}
\includegraphics[width=\textwidth]{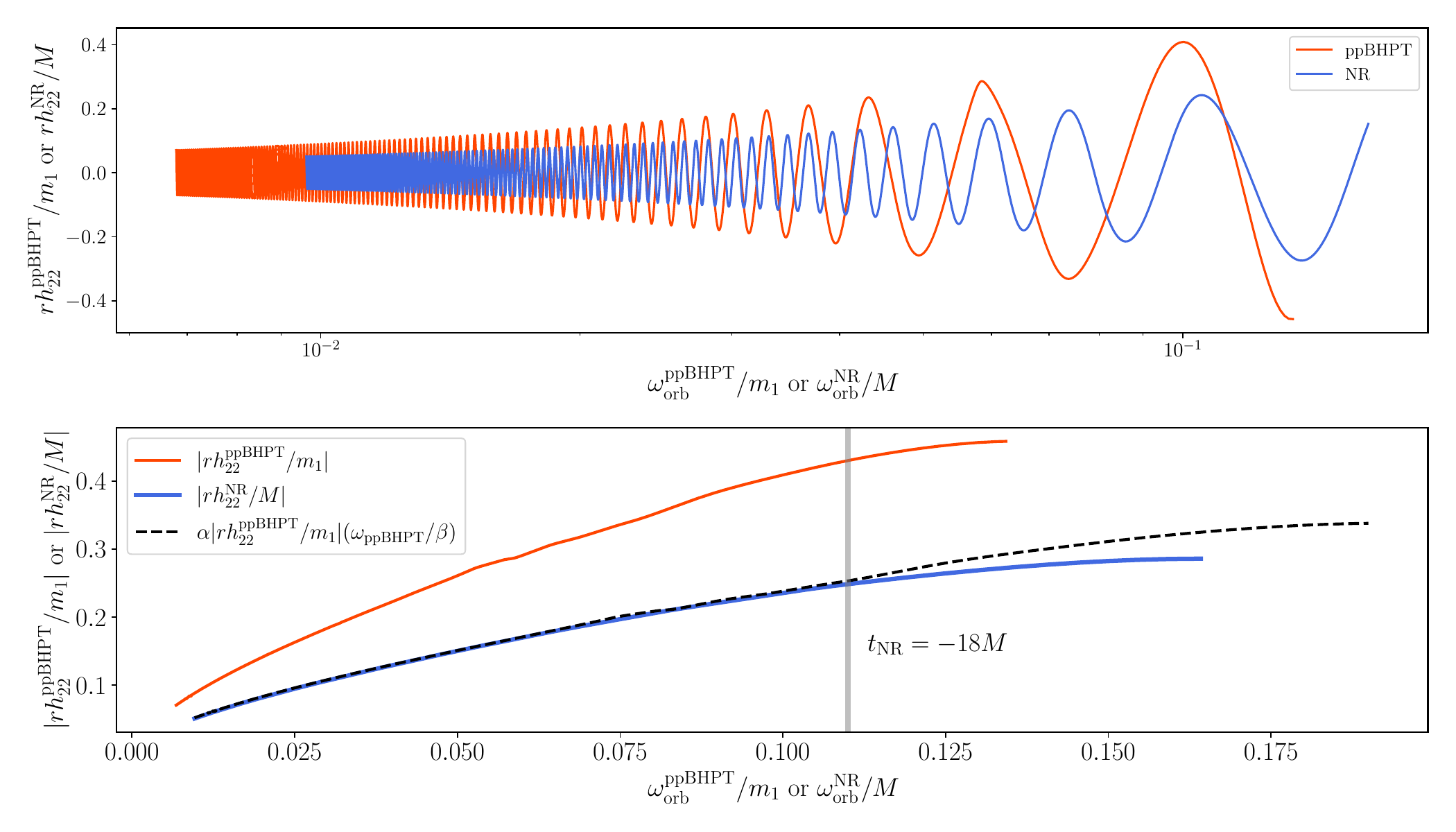}
\caption{We show the $(2,2)$ mode of the NR and ppBHPT waveforms, up to the merger i.e. $t=0$, as a function of the orbital frequencies, using the mass-scale of $M$ for NR and $m_1$ for ppBHPT (upper panel) for $q=3$. The lower panel shows the amplitudes of the waveforms as a function of the orbital frequencies. For comparison, we also show the amplitudes obtained using the approximate scaling in Eq.(\ref{eq:alpha_beta_freq}) (lower panel). The grey vertical line represents the time $t_{\rm NR}=-18M$ upto which the scaling works really well. More details are in Section \ref{sec:alpha_beta_freq}.}
\label{fig:alpha_beta_func_of_frequencies}
\end{figure*}

In Figure~\ref{fig:q3_optimal_alpha_beta}, we compare the obtained values of $\alpha$ and $\beta$ from different approaches. We observe that $\alpha_{\rm peak}$ remains relatively constant throughout the binary evolution, while $\beta_{\rm peak}$ shows stability in the earlier stages and deviates slightly during the late-inspiral-merger phase. It is important to note that $\alpha_{\rm peak}$ and $\beta_{\rm peak}$ represent local optimal values and may differ slightly from the global fit values, e.g. $\alpha_{5000M}$ and $\beta_{5000M}$. We also examine $\alpha_{\rm 5cycles}$ and $\beta_{\rm 5cycles}$, which provide averaged local estimations of the calibration parameters. We note that $\alpha_{\rm 5cycles}$ closely follows $\alpha_{\rm peak}$, while $\beta_{\rm 5cycles}$ aligns well with $\beta_{\rm peak}$, except for the late-inspiral and merger region where some deviations occur for $\beta$. We further note that the obtained values of $\alpha$ and $\beta$ from the different approaches are not simply consistent with the naive mass-scale transformation of $\frac{1}{1+1/q}$. This naive mass-scale transformation is required to transform the mass-scale of the ppBHPT waveforms from $m_1$ to $M$.
This suggests that the calibration parameters $\alpha$ and $\beta$ encompass additional effects beyond a simple mass-scale transformation. Next, we plot the $\alpha$ and $\beta$ from different approaches as a function of the NR orbital frequencies (Figure~\ref{fig:q3_optimal_alpha_beta_freq}). This further demonstrates that the $\alpha$ and $\beta$ values are mostly constant for a significant portion of the frequency window. 

\begin{figure*}
\includegraphics[scale=0.47]{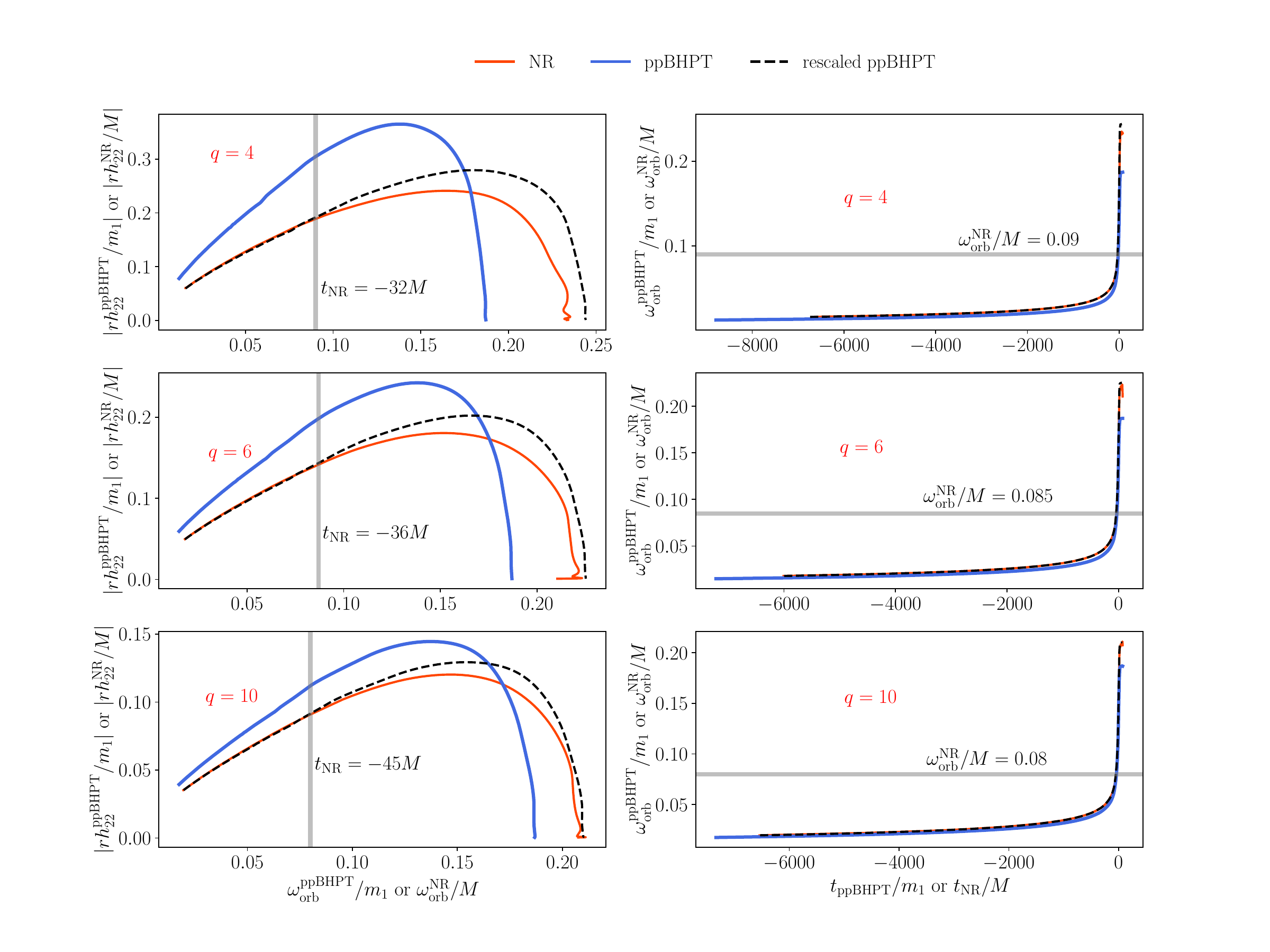}
\caption{We show the amplitudes of ppBHPT (red lines) and NR waveforms (blue lines) as a function of orbital frequencies (left panels) for $q=[4,6,10]$. Additionally, we include the amplitudes obtained from the approximate scaling given by Eq.(\ref{eq:alpha_beta_freq}) (black dashed lines) for comparison. In right panels, we show the orbital frequencies as a function of time. The mass-scale of NR and ppBHPT waveforms are $M$ and $m_1$ respectively. In the left panels, grey vertical lines indicate the time $t_{\rm NR}$ up to which the scaling is effective, while in the right panels, grey horizontal lines represent the orbital frequency $\omega_{\rm orb}$ up to which the scaling holds. Further details can be found in Section \ref{sec:alpha_beta_freq}.}
\label{fig:q_4_6_10}
\end{figure*}

In Figure~\ref{fig:q3_alpha_beta_optimization}, we therefore investigate the applicability of the $\alpha$-$\beta$ scaling to the entire length of the available NR data by utilizing the full $30000M$ of NR waveform data, covering 56 cycles, along with the corresponding $\sim35000m_1$ ppBHPT waveform data. By employing Eq.(\ref{eq:EMRI_rescale}) and following the procedure outlined in Sec.\ref{sec:scaling_full_wf}, we successfully obtain a set of $\alpha$ and $\beta$ values that allow us to rescale the full ppBHPT waveform to match the NR data throughout the binary evolution. Note that these values are denoted as $\alpha_{\rm full}$ and $\beta_{\rm full}$ and are also shown in Figure.~\ref{fig:q3_optimal_alpha_beta} and ~\ref{fig:q3_optimal_alpha_beta_freq} for comparison. In the top row of Figure~\ref{fig:q3_alpha_beta_optimization}, we show the NR data and the ppBHPT waveforms after applying the scaling factor of $\frac{1}{1+1/q}$. Additionally, we present the rescaled ppBHPT waveform after the $\alpha$-$\beta$ calibration in the second row. In the third row of Figure~\ref{fig:q3_alpha_beta_optimization}, we show $\Delta A/A_{\rm NR}$, relative error in amplitude, and $\Delta \phi_{\rm NR}$, absolute error in the phase, of both the ppBHPT (after multiplying the waveform with the factor $\frac{1}{1+1/q}$ to have the same mass-scale of NR) and rescaled ppBHPT waveform when compared to the NR data. These errors indicate that the rescaled ppBHPT waveform exhibits excellent agreement with the NR data, with amplitude errors on the order of $\sim 0.1$\% and phase errors of approximately $\sim 0.1$ radians. These errors are significantly smaller compared to the errors between the original ppBHPT waveform (after multiplying the factor of $\frac{1}{1+1/q}$) and the NR data, demonstrating the effectiveness of the $\alpha$-$\beta$ scaling in improving the agreement between the two waveforms.

Finally, to understand and mitigate the effect of the merger-ringdown waveform in the $\alpha$-$\beta$ calibration, we follow the procedure outlined in Section~\ref{sec:scaling_inspiral_wf} and use only the waveform up to $t=-100M$. The resulting calibration parameters are denoted as $\alpha_{\rm ins}$ and $\beta_{\rm ins}$. We find that $\alpha_{\rm ins}$ and $\beta_{\rm ins}$ are very close to $\alpha_{\rm full}$ and $\beta_{\rm full}$, respectively. Specifically, we have $$[\alpha_{\rm full},\beta_{\rm full}]=[0.737122,0.706900]$$ and $$[\alpha_{\rm ins},\beta_{\rm ins}]=[0.731040,0.707100].$$ We show the values in Figure.~\ref{fig:q3_optimal_alpha_beta} and ~\ref{fig:q3_optimal_alpha_beta_freq} for comparison. These values suggest that the inspiral-only waveform has a slightly larger effect on the $\alpha$ value compared to the $\beta$ value. However, since the values are very close, it implies that we can use any segment of the waveform and still obtain meaningful estimates for the $\alpha$ and $\beta$ parameters. We will demonstrate this later in Section~\ref{sec:PN_alpha_beta} (cf. Figs.~\ref{fig:q3_alpha_beta_optimization_PN} and~\ref{fig:alpha_beta_PN_NR}).

\subsection{Validity of the $\alpha$-$\beta$ scaling}
\label{sec:alpha_beta_validity}
The results presented in Section~\ref{sec:alpha_beta_comparison} provide valuable insights into the validity and behavior of the $\alpha$-$\beta$ scaling between ppBHPT waveforms and NR dat for $q=3$. This insight provides a reasonable understanding into the validity of the scaling in the comparable mass regime. The key findings are as follows:
\begin{itemize}
    \item The scaling procedure is effective even for longer NR simulations with a duration of approximately $30000M$. This demonstrates that the $\alpha$-$\beta$ scaling can be successfully applied to a wide range of waveform data, including those with a significant number of orbital cycles. 
    \item Throughout most of the binary evolution, the optimal values of $\alpha$ and $\beta$ remain approximately constant. This indicates that a global set of calibration parameters can reasonably capture the local behaviour.
    \item In the late-inspiral and merger stage, slight deviations from constant values are observed for both $\alpha$ and $\beta$. As a result, the scaling remains extremely effective until very close to merger (up to $\sim40M$ before the merger) beyond which slight differences between rescaled ppBHPT and NR is observed. We can attribute these deviations to the changes in mass and spin of the final black hole during this phase. Ref.~\cite{Islam:2023mob} has shown that the $\alpha$ and $\beta$ values, obtained in the inspiral part of the waveform, can be self-consistently rescaled for the merger-ringdown part using the energy and angular momenta changes up to plunge. In particular, the $\alpha_{\rm MR}$ and $\beta_{\rm MR}$, calibration parameters to match ppBHPT waveform to NR at the merger-ringdown part, obeys the following scaling with $\alpha_{\rm full}$ and $\beta_{\rm full}$~\cite{Islam:2023mob}:
    \begin{align} \label{eq:xi_def}
        \alpha_{\rm MR} = \hspace{1mm}\xi \times \alpha_{\rm full} \;,
    \end{align}
    and
    \begin{align} \label{eq:xi_def}
        \beta_{\rm MR} = \hspace{1mm}\frac{\beta_{\rm full}}{\xi} \; ,
    \end{align}
    where the scaling factor can be approximated as $\xi=[1-(\frac{\Delta J^z}{M^2})^{1.5}] (1-\frac{\Delta E}{M})$. Here, $\Delta E$ and $\Delta J^z$ are the change in energy and angular momentum up to the plunge. Additionally, $\alpha_{\rm MR}$ and $\beta_{\rm MR}$ are the scaling parameters at the merger-ringdown part.
    
\end{itemize}
Overall, these findings support the applicability and robustness of the $\alpha$-$\beta$ scaling approach in relating ppBHPT waveforms to NR data in the comparable mass regime.

\subsection{Understanding $\alpha$-$\beta$ scaling as frequency-dependent corrections}
\label{sec:alpha_beta_freq}
The $\alpha$-$\beta$ scaling between ppBHPT and NR waveforms is designed to address the missing finite size effects~\cite{Islam:2023aec} and higher-order self-force corrections~\cite{Wardell:2021fyy} in ppBHPT waveforms. Wardell et al. (Ref.~\cite{Wardell:2021fyy}) have shown that the second-order self-force correction (as well as the leading order term) is frequency dependent. It raises the question of how the $\alpha$-$\beta$ scaling can handle frequency-dependent corrections. We now therefore derive the $\alpha$-$\beta$ scaling as a function of the orbital frequencies and show explicitly that $\alpha$-$\beta$ scaling introduces frequency dependent corrections.

\begin{figure*}
\includegraphics[scale=0.47]{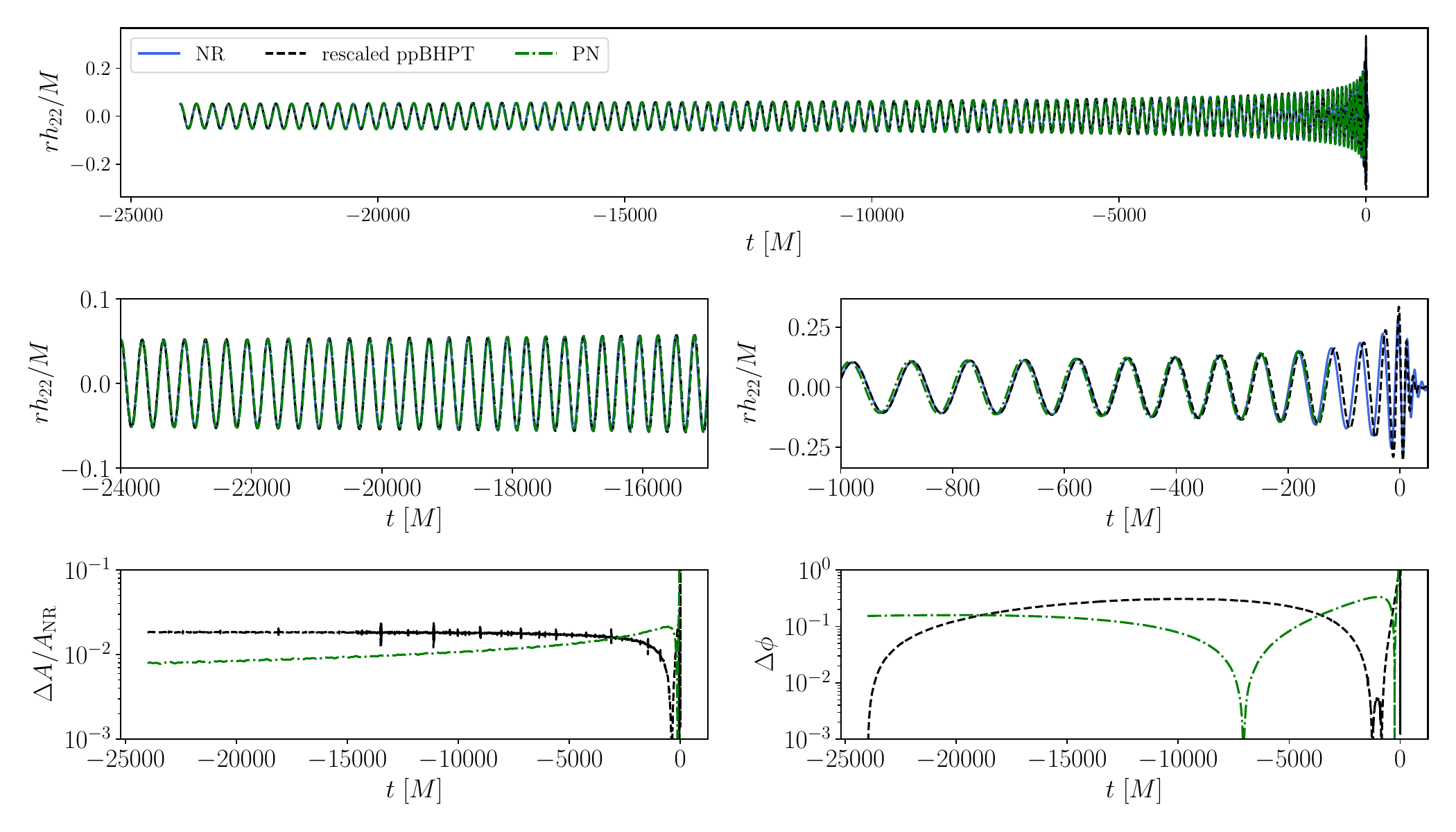}
\caption{We show the waveforms for the full $(2,2)$ mode inspiral-merger-ringdown from NR (blue solid lines), rescaled ppBHPT (black dashed lines), and PN (green dash-dotted lines) for $q=3$. All waveforms have the mass-scale of $M$. The upper panel displays the waveforms for the entire duration, while the second row focuses on the early inspiral and the merger-ringdown phases. In the third row of the figure, we provide the errors in both the rescaled ppBHPT and PN waveforms when compared to NR. Further details can be found in Section \ref{Sec:PN}.}
\label{fig:PN}
\end{figure*}

The $\alpha$-$\beta$ scaling for the $(2,2)$ mode can be expressed as follows:
\begin{align} \label{eq:EMRI_rescale_22}
h^{2,2}_{\tt NR}(t_{\tt NR} ; q) \sim \alpha h^{2,2}_{\tt ppBHPT}\left( \beta \times t_{\tt ppBHPT};q \right) \,.
\end{align}
This scaling relationship extends to the amplitude and phase of the waveforms:
\begin{align}
A^{2,2}_{\tt NR} (t_{\tt NR}) \approx \alpha \times A^{2,2}_{\tt ppBHPT}(\beta \times t_{\tt ppBHPT}),
\end{align}
and
\begin{align}
\phi^{2,2}_{\tt NR} (t_{\tt NR}) \approx \phi^{2,2}_{\tt ppBHPT}(\beta \times t_{\tt ppBHPT}).
\end{align}
One can compute the orbital phase as:
\begin{align}
\phi_{\tt orb, NR} = \phi^{2,2}_{\tt NR} / 2,\notag \\
\phi_{\tt orb, ppBHPT} = \phi^{2,2}_{\tt ppBHPT} / 2.
\end{align}
This leads to:
\begin{equation}
\frac{d \phi_{\tt orb, NR}}{dt_{\tt NR}} \approx \frac{d\phi_{\tt orb, ppBHPT}}{d(t_{\tt ppBHPT})} \frac{dt_{\tt ppBHPT}}{dt_{\tt NR}}.
\end{equation}
Simplifying further, we find:
\begin{equation}
\omega_{\tt orb, NR} \approx \omega_{\tt orb, ppBHPT} \frac{dt_{\tt ppBHPT}}{dt_{\tt NR}},
\end{equation}
where $\omega_{\tt NR}$ and $\omega_{\tt ppBHPT}$ are the orbital frequencies of the NR and ppBHPT waveforms respectively. Since $t_{\tt NR} = \beta t_{\tt ppBHPT}$, we can further simplify it as:
\begin{equation}
\omega_{\tt orb, NR} \approx \omega_{\tt orb, ppBHPT} \times \frac{1}{\beta}.
\end{equation}
Thus, the $\alpha$-$\beta$ scaling relationship between ppBHPT and NR waveforms (where both of them are expressed as a function of time), given in Eq.(\ref{eq:EMRI_rescale_22}), can be equivalently expressed as a scaling between the waveforms as a function of the orbital frequencies. These scalings are:
\begin{equation}
A^{2,2}_{\tt NR} (\omega_{\tt orb, NR}) \approx \alpha \times A^{2,2}_{\tt ppBHPT}(\frac{\omega_{\tt orb, ppBHPT}}{\beta}),
\end{equation}
and
\begin{equation}\label{eq:alpha_beta_freq}
h^{2,2}_{\tt NR} (\omega_{\tt orb, NR}) \approx \alpha \times h^{2,2}_{\rm ppBHPT}(\frac{\omega_{\tt orb, ppBHPT}}{\beta}).
\end{equation}

In Figure.~\ref{fig:alpha_beta_func_of_frequencies}, we present the $(2,2)$ mode of the NR and ppBHPT waveforms up to merger as a function of the orbital frequencies in the upper panel, accompanied by the amplitudes in the lower panel. It is evident that any rescaling aiming to match the ppBHPT amplitude (plotted against orbital frequencies; red solid line) to NR (also against orbital frequencies; blue solid line) must be frequency-dependent. We demonstrate that the $\alpha$-$\beta$ scaling described by Eq.(\ref{eq:EMRI_rescale_22}) corresponds to a frequency-dependent correction, as it not only modifies the amplitudes but also alters the frequency evolution according to Eq.(\ref{eq:alpha_beta_freq}). For comparison, we include the amplitudes as a function of rescaled frequencies (black dashed line) after the application of the $\alpha$-$\beta$ scaling. We find visual agreement up to $t_{\rm NR}=-18M$, very close to the merger, between the rescaled ppBHPT amplitudes as a function of rescaled orbital frequencies and NR amplitudes as a function of NR orbital frequencies.

Finally, we generalize the scaling for all modes as:
\begin{equation}\label{eq:alpha_beta_freq}
h^{\ell,m}_{\tt NR} (\omega_{\tt orb, NR}) \approx \alpha_{\ell} \times h^{\ell,m}_{\tt ppBHPT}\left(\frac{\omega_{\tt orb, ppBHPT}}{\beta}\right).
\end{equation}

To further support our observations, we extend our analysis to three additional mass ratio values: $q=[4,6,10]$, using publicly available SXS NR data \texttt{SXS:BBH:1220}~\cite{sxs_1220}, \texttt{SXS:BBH:0181}~\cite{sxs_0181}, and \texttt{SXS:BBH:1107}~\cite{sxs_1107}, respectively. However, these NR datasets only cover the final $\sim 6000M$ evolution of the binary, corresponding to approximately 25 orbital cycles. For each mass ratio, we perform the $\alpha$-$\beta$ scaling using Eq.(\ref{eq:EMRI_rescale_22}), obtaining the best-fit values for $\alpha$ and $\beta$. We then use Eq.(\ref{eq:alpha_beta_freq}) to approximate the rescaled amplitude as a function of the orbital frequency (Fig.~\ref{fig:q_4_6_10}).

In Figure~\ref{fig:q_4_6_10} (left panels), we compare the amplitudes of both ppBHPT and NR waveforms as a function of the respective orbital frequencies. To gain better understanding, we also plot the orbital frequencies as a function of time in the right panels. During the inspiral phase, the rescaled waveform's amplitude closely matches NR, but deviations become apparent as it approaches the merger. However, the approximate $\alpha$-$\beta$ scaling effectively captures the frequency-dependent correction needed to align ppBHPT with NR until very close to the merger, where the scaling breaks down. Specifically, we find that the scaling remains effective up to $t_{\rm NR}=-32M$ for $q=4$, $t_{\rm NR}=-36M$ for $q=6$, and $t_{\rm NR}=-45M$ for $q=10$. This suggests that the $\alpha$-$\beta$ scaling successfully matches NR data very well up to the plunge phase. 

It is worth mentioning that the reason for the global $\alpha$-$\beta$ fit to be less effective around merger is that the global fit values deviate from the local optimal $\alpha$-$\beta$ estimates in this regime (Fig.~\ref{fig:q3_optimal_alpha_beta}). These deviations can also be attributed to the changes in mass and spin of the final black hole during this phase~\cite{Islam:2023mob}. Incorporating the updated final mass and spin values in the ppBHPT framework is expected to reduce these deviations and improve the accuracy of the rescaling~\cite{Islam:2023mob}.

\begin{figure}
\includegraphics[width=\columnwidth]{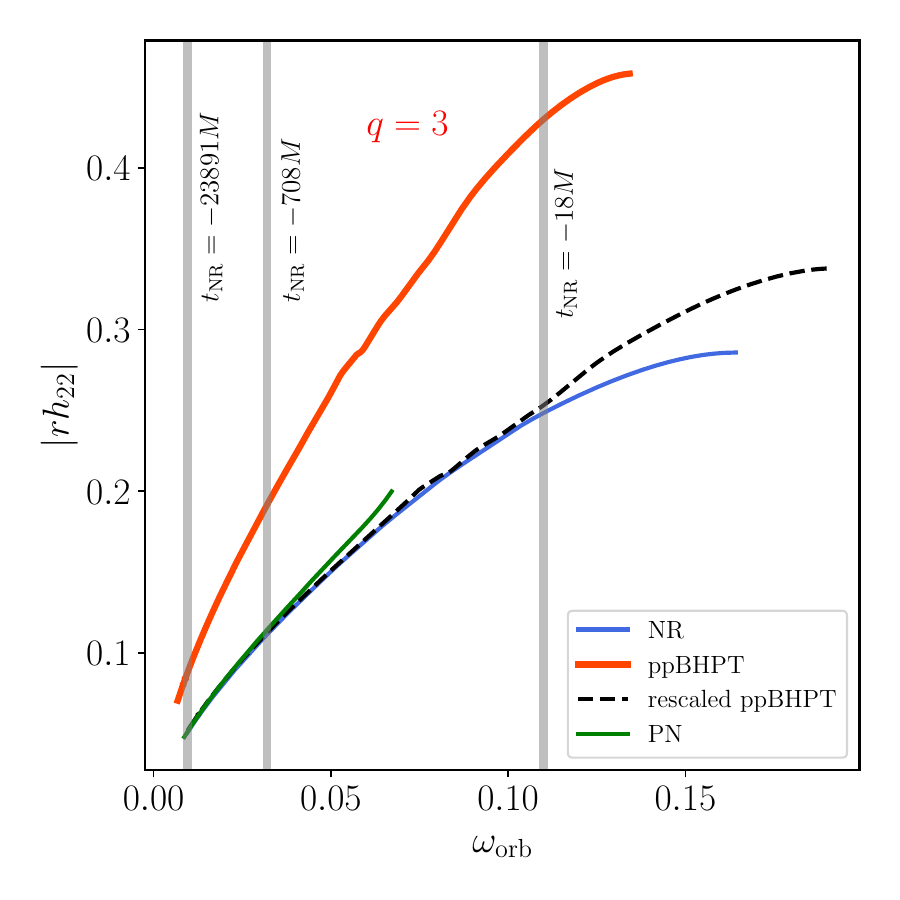}
\caption{We show the amplitudes of ppBHPT (red lines), NR waveforms (blue lines) and PN (green lines) as a function of orbital frequencies for $q=3$. Additionally, we include the amplitudes obtained from the approximate scaling given by Eq.(\ref{eq:alpha_beta_freq}) (black dashed lines) for comparison. The mass-scale of NR and ppBHPT waveforms are $M$ and $m_1$ respectively. Grey vertical lines indicate the time $t_{\rm NR}$ up to which PN and $\alpha$-$\beta$ rescaling shows remarkable match with NR. More details are in Section~\ref{Sec:PN}.}
\label{fig:PN_amp_freq}
\end{figure}

\section{Comparison against Post-Newtonian Theory}
\label{Sec:PN}
We now provide a detailed comparison of the post-Newtonian theory waveforms with ppBHPT, rescaled ppBHPT (obtained through the $\alpha$-$\beta$ procedure) and NR in the comparable mass regime. A detailed review of post-Newtonian methods are given in Ref.~\cite{Blanchet:2013haa}. The post-Newtonian approximation is a slow-motion,
weak-field approximation to general relativity with an expansion parameter $\xi = \frac{v}{c}$ where $v$ is the magnitude of the relative velocity and $c$ is the speed of light. While many previous analysis have focused on understanding the match between NR and PN in the comparable mass regime~\cite{Boyle:2007ft,Hannam:2007wf,Pan:2007nw,Hannam:2007ik}, our focus remains in comparing ppBHPT to PN.

\subsection{Comparing waveforms at $q=3$}
\label{sec:PN_wf}
We show the full $(2,2)$ mode inspiral-merger-ringdown waveforms from NR (blue solid lines), rescaled ppBHPT (black dashed lines), and PN (green dash-dotted lines) in Fig.~\ref{fig:PN}. In particular, we use \texttt{TaylorT4} PN approximation, generated using \texttt{LALSimulation} software package. This particular approximation includes phase terms up to 3.5PN order and amplitude terms up to 2.5PN order~\cite{Boyle:2007ft}. We zoom into the earlier and later times of the waveform to examine the match between rescaled ppBHPT, PN, and NR in more detail. Additionally, we compute the relative error in the amplitude and the absolute phase error for both rescaled ppBHPT and PN compared to NR. The results indicate that both rescaled ppBHPT and PN exhibit similar errors in the amplitude when compared to NR. However, in the late inspiral phase (e.g. $-4000M \leq t \leq -100M$), the rescaled ppBHPT waveform yields a much smaller error in the phase compared to the PN waveform. This suggests that the rescaled ppBHPT waveform provides improved accuracy in capturing the phase evolution of the system during the late inspiral regime, as compared to the PN approximation. This is expected as, in the late inspiral, the binary moves into the strong field and the PN approximation breaks down.

\begin{figure*}
\includegraphics[width=\textwidth]{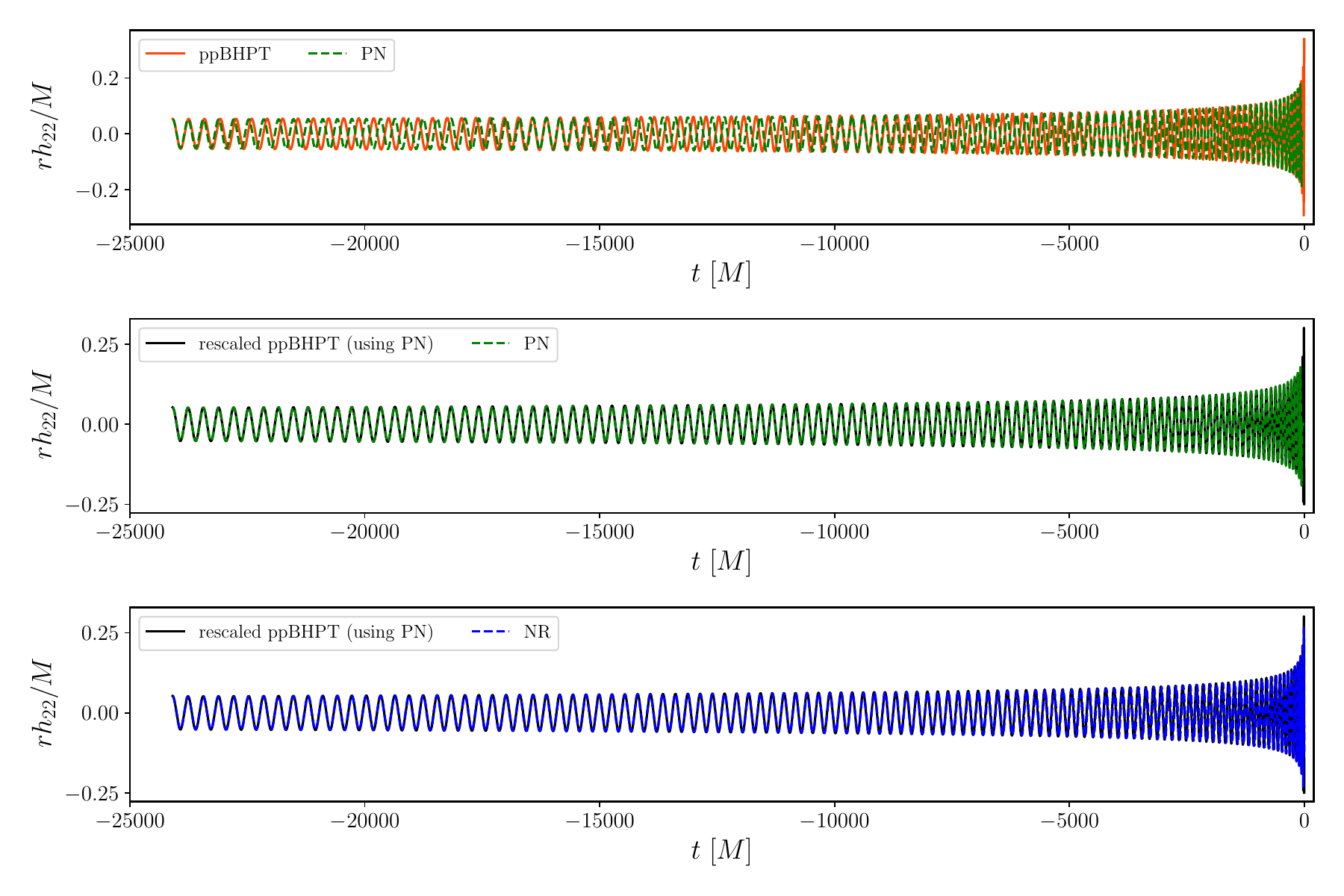}
\caption{We show the comparison of the $(2,2)$ mode of the PN  and ppBHPT waveforms (first row), along with the comparison between PN and rescaled ppBHPT waveforms (second row) for $q=3$. Rescaled ppBHPT waveform is obtained by performing an $\alpha$-$\beta$ calibration to PN. For comparison, we also show the NR waveform in the third row. All waveforms have the mass-scale of $M$. More details are in Section \ref{sec:PN_alpha_beta}.}
\label{fig:q3_alpha_beta_optimization_PN}
\end{figure*}

Finally, in Figure.~\ref{fig:PN_amp_freq}, we present the amplitude of the ppBHPT, rescaled ppBHPT, NR, and PN waveforms as a function of the respective orbital frequencies. The amplitude evolution extracted from NR is compared to both the PN and rescaled ppBHPT waveforms. We observe that in the inspiral region, both the PN and rescaled ppBHPT amplitudes agree well with the amplitude evolution obtained from NR. However, as we progress towards later times, the PN approximation starts deviating from NR around $t_{\rm NR}=-708M$, while the rescaled ppBHPT approximation breaks down around $t_{\rm NR}=-18M$. This suggests that the rescaled ppBHPT waveform better captures the dynamics of NR compared to the PN waveform.

\subsection{Estimating $\alpha$-$\beta$ using PN}
\label{sec:PN_alpha_beta}
Our analysis in Section~\ref{sec:alpha_beta_comparison}, Section~\ref{sec:alpha_beta_validity}, and Section~\ref{sec:PN_wf} highlights interesting possibilities for using PN waveforms to accurately estimate global values for $\alpha$ and $\beta$. We are motivated by the following observations:
\begin{itemize}
    \item The values of $\alpha$ and $\beta$ remain nearly constant for a significant duration of the binary evolution, with only slight deviations around the merger. These local estimates of $\alpha$ and $\beta$ closely align with the values obtained using global error minimization techniques. Furthermore, the values of $\alpha$ and $\beta$ obtained using only the inspiral part of the waveform exhibit remarkable agreement with the values obtained using the full waveform data.
    \item While the PN approximation breaks down towards the merger (e.g. at $t=-708M$ in Fig.~\ref{fig:PN_amp_freq}; or about 11 cycles before merger as reported in Ref.~\cite{Hannam:2007ik}), it provides an excellent match to NR waveforms in the inspiral phase, which is far away from the merger.
\end{itemize}
These observations suggest that PN waveforms can be used to infer the $\alpha$ and $\beta$ values required to match a ppBHPT waveform to PN. These $\alpha$ and $\beta$ values will be very close to the values obtained using NR data. This procedure has significant implications. Firstly, it means that one can use ppBHPT and PN waveforms in the early inspiral to obtain the $\alpha$-$\beta$ values and generate a rescaled ppBHPT waveform that matches NR waveforms throughout the entire binary evolution, from inspiral to ringdown.

In this section, we investigate the possibility of using PN waveforms for estimating $\alpha$ and $\beta$ in great detail using different PN approximations.

\subsubsection{$\alpha$-$\beta$ PN scaling at $q=3$}
\label{sec:PN_alpha_beta_q3}
First, we perform a calibration between the ppBHPT waveform at $q=3$ and a PN waveform generated using the \texttt{TaylorT4} approximation. We obtain $\alpha_{\rm PN}$ and $\beta_{\rm PN}$ as the calibration parameters. Interestingly, we find that these values are very close to $\alpha_{\rm NR, ins}$ and $\beta_{\rm iNR, ins}$ obtained by comparing the inspiral portion of the NR and ppBHPT waveforms, as well as $\alpha_{\rm NR, full}$ and $\beta_{\rm NR, full}$ obtained from the comparison of full NR and ppBHPT waveforms. Specifically, we have:
\begin{align*}
[\alpha_{\rm PN},\beta_{\rm PN}] &=[0.738862,0.705607], \\
[\alpha_{\rm NR, full},\beta_{\rm NR, full}] &=[0.737122,0.706900], \\
[\alpha_{\rm NR, ins},\beta_{\rm iNR, ins}] &=[0.731040,0.707100].
\end{align*}
Furthermore, we utilize $\alpha_{\rm PN}$ and $\beta_{\rm PN}$ to rescale the ppBHPT waveform, and we observe an excellent match with the NR data not only in the inspiral phase (Figure.~\ref{fig:q3_alpha_beta_optimization_PN}, third row). We however notice some dephasing in the merger-ringdown part. Nonetheless, this analysis suggests that PN waveforms, which mostly capture the inspiral phase, can provide meaningful estimates of $\alpha$ and $\beta$ for rescaling ppBHPT waveforms to match NR waveforms reasonably well in the inspiral part. For example, the $L_2$-norm error between ppBHPT and NR waveform up to merger in Fig.~\ref{fig:q3_alpha_beta_optimization_PN} is $\sim0.9$. However, the $L_2$-norm error between PN and NR in that time window is $\sim0.02$ whereas the error between PN-rescaled ppBHPT and NR is $\sim0.06$. Once we obtain the scaled ppBHPT waveforms for the inspiral, we can then utilize the framework developed in Ref.~\cite{Islam:2023mob} to obtain appropriately scaled ppBHPT waveform at the merger-ringdown part too.

\subsubsection{$\alpha$-$\beta$ PN scaling at $q=[4,6,10]$}
\label{sec:PN_alpha_beta_q34_6_10}
\begin{figure}
\includegraphics[width=\columnwidth]{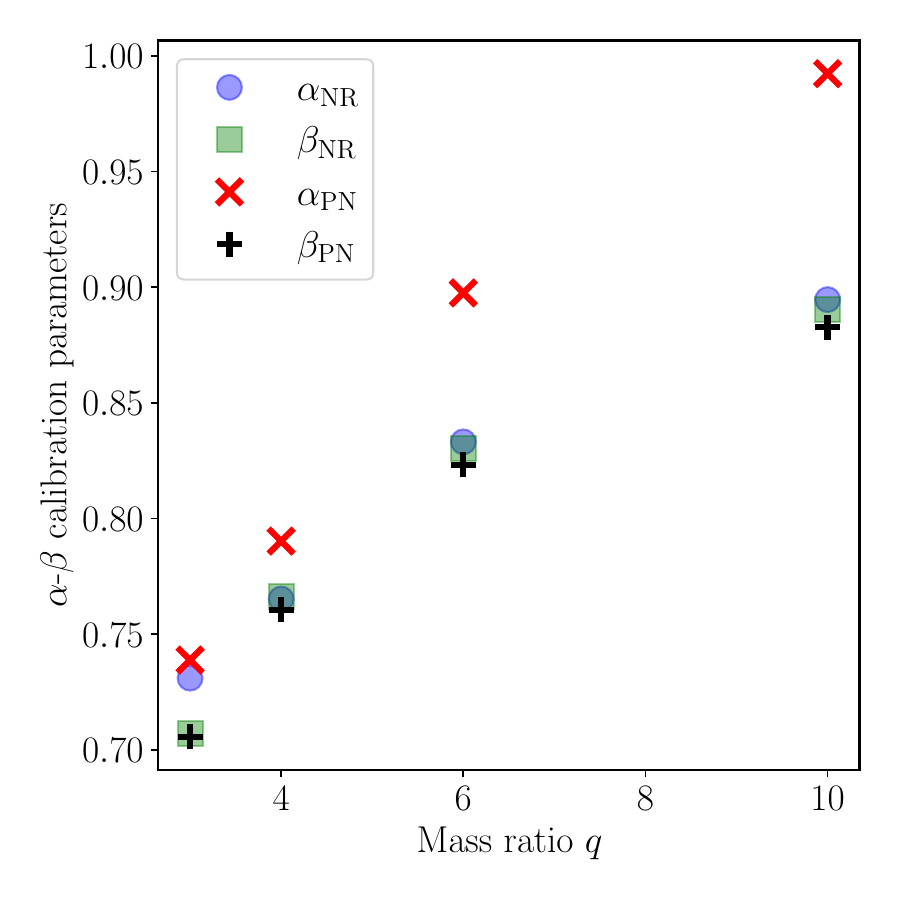}
\caption{We show the $\alpha$ and $\beta$ values for a set of mass ratios obtained by performing the calibration against NR and PN waveforms. For PN, we use only the inspiral data to obtain the values of $\alpha$ and $\beta$ whereas full IMR waveform is used for NR. More details are in Section \ref{sec:PN_alpha_beta_q34_6_10}.}
\label{fig:alpha_beta_PN_NR}
\end{figure}

To investigate the validity of our observations for different mass ratios, we repeat the analysis for mass ratios $q=4$, $q=6$, and $q=10$. In Figure~\ref{fig:alpha_beta_PN_NR}, we present the values of $\alpha$ and $\beta$ obtained by rescaling the ppBHPT waveforms to both NR and PN data. We find that $\beta_{\rm PN}$, obtained from the PN waveform, closely matches $\beta_{\rm NR}$ for all mass ratios. This suggests that the $\beta$ parameter is relatively insensitive to the choice of waveform and is consistent between NR and PN. However, we observe that $\alpha_{\rm PN}$, also obtained from the PN waveform, is systematically larger than the values obtained from NR. This difference appears to increase as the mass ratio increases. Nevertheless, it is noteworthy that the values of $\alpha$ obtained from PN are still quite close to those obtained from NR, indicating a reasonable agreement between the two.

\subsubsection{Understanding the effect of the choice of PN model}
\label{sec:PN_alpha_beta_TaylorT1234}
It is important to note that each PN model includes corrections up to a certain PN order, and these higher-order corrections can affect the accuracy of the rescaling. To investigate the effect of different PN models on the $\alpha$-$\beta$ calibration, we repeat the calibration process for $q=3$ using different PN approximations: \texttt{TaylorT1}, \texttt{TaylorT2}, and \texttt{TaylorT4}. While all of these approximation includes phase terms up to 3.5PN order and amplitude terms up to 2.5PN order, they employ different techniques and expansions to obtain these terms~\cite{Damour:2000zb,Damour:2002kr}. This allows us to assess whether the choice of PN model affects the resulting values of $\alpha$ and $\beta$. By performing the $\alpha$-$\beta$ calibration with different PN approximations, we obtain slightly different values for $\alpha$ and $\beta$. In particular, we find:
\begin{align*}
[\alpha_{\rm PN}^{\rm TaylorT4},\beta_{\rm PN}^{\rm TaylorT4}] &=[0.738862,0.705607], \\
[\alpha_{\rm PN}^{\rm TaylorT1},\beta_{\rm PN}^{\rm TaylorT1}] &=[0.745278,0.709454], \\
[\alpha_{\rm PN}^{\rm TaylorT2},\beta_{\rm PN}^{\rm TaylorT2}] &=[0.753860,0.709265].
\end{align*}
It is interesting to note that value of $\beta$ changes marginally when we use a different PN model. However, changes in $\alpha$ is more prominent. This indicates that the choice of PN model does have a slight an impact on the rescaling parameters.

\section{Discussion \& Conclusion}
\label{Sec:Discussion}
In this paper, we investigated the validity and effectiveness of the $\alpha$-$\beta$ scaling approach, previously introduced by Islam \textit{et al.}~\cite{Islam:2022laz}, which aims to match the ppBHPT waveforms to the NR waveforms. Utilizing publicly available long NR data (\texttt{SXS:BBH:2265}) for mass ratio $q=3$, we demonstrated that the scaling can be achieved even for longer NR simulations, spanning up to $\sim 30000M$ in duration. Throughout most of the binary evolution, the scaling factors $\alpha$ and $\beta$ can be computed utilizing publicly available long NR data (\texttt{SXS:BBH:2265}) for mass ratio $q=3$ and considered approximately constant, although they show slight deviations close to the merger. These deviations are expected due to the loss of energy and change in mass and spin of the final black hole during the merger process. Once the final mass and spin values are incorporated into the ppBHPT framework, these deviations are expected to be reduced~\cite{Islam:2023mob}.

Furthermore, we investigated the frequency-dependent nature of the scaling. We derived the $\alpha$-$\beta$ scaling as a function of orbital frequencies and demonstrated its equivalence to a frequency-dependent correction. The rescaled ppBHPT waveform, when matched to NR amplitudes as a function of orbital frequencies, showed excellent agreement, providing further support for the frequency-dependent nature of the scaling.

We then compared the accuracy of the rescaled ppBHPT waveform obtained through the $\alpha$-$\beta$ scaling against the \texttt{TaylorT4} post-Newtonian (PN) approximation. The rescaled ppBHPT waveform showed comparable accuracy to the PN waveform in terms of amplitude, but exhibited significantly smaller phase errors during the late inspiral phase.
Our analysis confirms the feasibility of using PN waveforms to derive precise $\alpha$-$\beta$ calibration parameters. The calibration process involves matching the ppBHPT waveform to a PN waveform, focusing on the inspiral phase. The resulting $\alpha$ and $\beta$ values obtained from this calibration closely align with those obtained from NR waveforms.

Overall, our results demonstrate that the $\alpha$-$\beta$ scaling provides an effective method for matching ppBHPT waveforms to NR waveforms in the comparable mass regime, accounting for missing finite-size effects and possibly higher-order self-force corrections~\cite{Islam:2023aec,Islam:2023qyt}. The scaling is frequency-dependent, capturing the correct amplitude and frequency evolution of the NR waveforms. While the scaling has limitations close to the merger (due to a change in mass and spin values~\cite{Islam:2023mob}), it remains highly effective in reproducing NR dynamics up to the plunge phase. These findings have implications for gravitational wave observations and waveform modeling in extreme-mass-ratio inspirals.

\begin{acknowledgments}
We thank Scott Field, Scott Hughes, Adam Pound, Niels Warburton, Barry Wardell and Chandra Kant Mishra for helpful discussions and thoughtful comments on the manuscript.
The authors acknowledge support of NSF Grants PHY-2106755, PHY-2307236 (G.K) and DMS-1912716, DMS-2309609 (T.I and G.K).  Simulations were performed on CARNiE at the Center for Scientific Computing and Visualization Research (CSCVR) of UMassD, which is supported by the ONR/DURIP Grant No.\ N00014181255 and the UMass-URI UNITY supercomputer supported by the Massachusetts Green High Performance Computing Center (MGHPCC). 
\end{acknowledgments}

\bibliography{References}

\end{document}